\documentclass[aps,prb,twocolumn,showpacs,floatfix,%
superscriptaddress,amssymb,amsmath]{revtex4}

\usepackage{graphicx}
\usepackage{dcolumn}
\usepackage{bm}

\begin{document}

\title{Phonon Rabi-assisted tunneling in diatomic molecules}

\author{E. Vernek}
\affiliation{Department of Physics and Astronomy, and Nanoscale
and Quantum Phenomena Institute, \\Ohio University, Athens, Ohio
45701-2979} \affiliation{Departamento de F\'{\i}sica,
Pontif\'{\i}cia Universidade Cat\'olica, Rio de Janeiro-RJ,
Brazil}

\author{E. V. Anda}
\affiliation{Departamento de F\'{\i}sica, Pontif\'{\i}cia
Universidade Cat\'olica, Rio de Janeiro-RJ, Brazil}

\author{S. E. Ulloa}
\affiliation{Department of Physics and Astronomy, and Nanoscale
and Quantum Phenomena Institute, \\Ohio University, Athens, Ohio
45701-2979}

\author{N. Sandler}
\affiliation{Department of Physics and Astronomy, and Nanoscale
and Quantum Phenomena Institute, \\Ohio University, Athens, Ohio
45701-2979}

\date{21 May 2005}

\begin{abstract}
We study electronic transport in diatomic molecules connected to
metallic contacts in the regime where {\em both} electron-electron
and electron-phonon interactions are important. We find that the
competition between these interactions results in unique resonant
conditions for interlevel transitions and polaron formation: the
Coulomb repulsion requires additional energy when electrons
attempt phonon-assisted interlevel jumps between fully or
partially occupied levels. We apply the equations of motion
approach to calculate the electronic Green's functions. The
density of states and conductance through the system are shown to
exhibit interesting Rabi-like splitting of Coulomb blockade peaks
and strong temperature dependence under the {\em interacting}
resonant conditions.
\end{abstract}


\maketitle

A significant current effort in nanoscopic systems is the study of
electron transport in natural and quantum-dot molecules. Much of
the interest lies in being able to investigate different regimes
of competing electron-electron and electron-phonon interactions.
It is typically the case, due to the spatial confinement, that
electron-electron interactions (EEI) play a more
important role than electron-phonon interactions (EPI) in
determining electronic transport properties in low-dimensional
systems.

Different geometries of quantum-dot molecules (constructed with
interconnected quantum dots) have been studied in the literature,
and the role of phonons on electron transport has been analyzed in
these systems. \cite{gen_QD_ref} For instance, it is known that
phonons are a relatively weak perturbation, responsible for the
broadening of Coulomb blockade peaks in the conductance and for
the appearance of satellite features in the nonlinear transport
regime. \cite{gen_QD_ref}

More recently, the field of ``molecular electronics,"
\cite{leo,james} where electrons and/or holes are injected
directly into molecules attached to metal electrodes, has seen
intense activity and progress. \cite{garcia,smit} It is
interesting to note that EPI become more important in molecular
electronics, since local molecule deformations produce significant
electronic level shifts, as has been observed in experiments.
\cite{Park} In fact, vibrational and torsional modes play
prominent roles in electron transport, producing sidebands in the
voltage-dependent differential conductance
\cite{McEuen,Flensberg,zhang} and/or polaronic shifts of the
electronic levels. \cite{Mitra} Furthermore, in a molecular system
with discrete electronic energy levels, vibrations produce
important effects when the energy of the vibrational modes matches
the energy difference between electronic levels.\cite{vasilevisky} As a result, EPI provides relaxation mechanisms (inelastic scattering) that affect
the conductance of the system. \cite{Orellana}

The simplest model of EPI is perhaps the independent boson model,
\cite{Mahan} where localized electrons interact with a phonon
system. Phonons introduce a shift of electronic levels and create
a series of phonon replica peaks in the density of states (
DOS) of the electronic system. In a double barrier
heterostructure interacting with phonons, as electrons have access
to an energy continuum in the region outside the barriers, the
inelastic scattering strongly affects the resonant tunneling
regime through the heterostructure. \cite{anda1} The effects of
EPI on a double-level quantum dot have also been studied recently,
\cite{eto} although the combined effects of inelastic scattering
and EEI were not considered.

\begin{figure}
\includegraphics{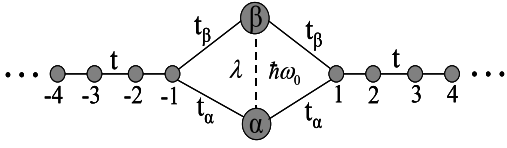}
 \caption{\label{Fig1} Schematic representation of the model system.
The electron-phonon interaction connects the local sites
($\epsilon_\alpha < \epsilon_\beta$) via phonons of frequency
$\omega_0$ with coupling constant $\lambda$.}
\end{figure}

In order to study a more realistic molecular system it is
important to include {\em both} interactions--- EEI and EPI ---
simultaneously. The interplay between the competing interactions
is likely to result in unique conditions for phonon emission and
absorption, as well as in unexpected polaron behavior.
\cite{Mitra} In this work we study the effect that EEI and EPI
have on charge transport through a diatomic molecule, envisioned
as two atomic sites (or quantum dots) directly coupled to leads,
as shown schematically in Fig.\ \ref{Fig1}.  We study this system
using the equations of motion method, which allows us to obtain
the DOS and electronic occupation as well as the conductance
through the system. We exploit the fact that the EPI strength
$\lambda$ is small compared to the phonon energy, and thus include
self-energy terms up to second order in this parameter. Standard
considerations, similar to those applied in the Hubbard
approximation, \cite{hubbard1,zubarev} are used to evaluate the
equations.  A significant result is the identification of 
occupation-dependent resonant conditions for phonon absorption and
emission in the presence of Coulomb repulsion.  More importantly,
we find a unique type of Rabi splitting in the DOS from the mixing
between a doubly occupied low-energy level (boosted by the Coulomb
repulsion) and a higher-energy state. This Rabi splitting is
mediated by thermally regulated phonon emission and absorption in
the molecule. The effect is shown to dramatically modify the
transport properties of the system since the resulting polaron
formation competes with resonant tunneling. The effect is made
remarkably noticeable even for weak EPI since the phonon-assisted
transport is magnified by the virtual emission and absorption
processes in the {\em interacting} resonant regime.

For concreteness we consider a two-level diatomic molecule with
local energies $\epsilon_{\alpha}^0$ and $\epsilon_{\beta}^0$ (we
assume $\epsilon_{\beta}^0 > \epsilon_{\alpha}^0$), as shown in
Fig.\ \ref{Fig1}.  Each dot or atomic site is connected
independently to two external current leads.  The system as
described can be mapped to that studied in Ref.~[\onlinecite{eto}]
(there for the one-electron case) and is designed to model various
experimental geometries. The total Hamiltonian is written as
$H_T=H_{mol}+H_{leads}+H_{mol-leads}$. Each lead is modeled as a
semi-infinite tight-binding chain, $H_{lead}=\sum_{\sigma,<j',j>}
tc^{\dagger}_{j'\sigma}c_{j\sigma}$, where the site index sum is
over nearest neighbors, $c^{\dagger}_{j\sigma}$ ($c_{j\sigma}$)
creates (annihilates) a fermion at the $j$-th site with spin $\sigma$,
and the left (right) lead is defined for $j',j \leq -1$ ($j',j
\geq 1$). The Hamiltonian for the molecule is given by
$H_{mol}=H_{el}+H_{ph}+H_{el-ph}$,
\begin{eqnarray}
H_{el}&=&\sum_{\sigma,i=\alpha,\beta}\left[\epsilon_i c^{\dagger}
_{i\sigma}c_{i\sigma} + \frac{U}{2} n_{i\sigma}
n_{i\bar{\sigma}}\right] \, , \\
H_{ph}&=&\left(b^{\dagger}b+\frac{1}{2}\right)\hbar\omega_0 \, ,
\end{eqnarray}
and in the rotating-wave approximation \cite{Mahan}
\begin{equation}
H_{el-ph}=\lambda\sum_{\sigma}\left(b^{\dagger}
c^{\dagger}_{\alpha\sigma}c_{\beta\sigma}+H.c\right) \, ,
\end{equation}
where $b^{\dagger}$ ($b$) creates (annihilates) a phonon with
energy $\hbar\omega_0$, $\epsilon_i=\epsilon^0_i-eV_g$, and $i =
\alpha, \beta$. The gate voltage $V_g$ controls the particle
number by shifting both localized energies with respect to the
Fermi energies of the left and right electrodes. We assume here
the same gate on both sites, as is likely the case in molecules.
\cite{Park} Finally, $H_{mol-leads}=\sum_{\sigma,i=\alpha,\beta \atop j=1,-1} t_i c^{\dagger}_{i\sigma} c_{j\sigma}+H.c.$, 
connects the diatomic molecule to the leads, where we will
consider $t_{\alpha},t_{\beta} \ll t$. Away from the Kondo regime
the effect of the leads is to broaden the energy levels of the
dots through the tunnel couplings $t_{\alpha}$ and $t_{\beta}$.
(Ref.~\onlinecite{f1})

To determine the dynamics of the electrons in the molecule, we
calculate the local retarded Green's function,
%
%
together with the nonlocal functions,
\begin{eqnarray}
iG^{\sigma}_{\alpha\beta\atop{\beta\alpha}}(\epsilon)
&=&\int \theta (t)\left[\langle
[bc_{\alpha\sigma}(t),c^{\dagger}_{\beta\sigma}(0)]_+
\rangle\atop{\langle
[b^{\dagger}c_{\beta\sigma}(t),c^{\dagger}_{\alpha\sigma}(0)]_
+\rangle}\right]e^{i\epsilon t}dt.
\end{eqnarray}
The latter are needed since they describe electron propagation due
to the EPI, and are associated with the absorption
($G^{\sigma}_{\alpha\beta}$) and emission
($G^{\sigma}_{\beta\alpha}$) of phonons. The corresponding
equations of motion up to second order in the EPI coupling
strength $\lambda$, yield ($i=\alpha, \beta$),
\begin{widetext}
\begin{eqnarray}
G_{ii}(\epsilon) &=&\frac{\epsilon-\epsilon_i-U\left(1-\langle
n_i\rangle\right)}{\left( \epsilon-\epsilon_i\right)\left(
\epsilon-\epsilon_i-U\right)-i\Gamma (\epsilon)\left[ \epsilon%
-\epsilon_i-U\left(1-\langle%
n_i\rangle\right)\right]-\Sigma_i(\epsilon)}\label{greenf},
\end{eqnarray}
\end{widetext}
\begin{eqnarray}
G_{\beta\alpha}(\epsilon)&=&\frac{\lambda\left[\langle
b^{\dagger}b\rangle +\langle n_{\beta}\rangle\right]
G_{\alpha\alpha}(\epsilon)}{\epsilon -\epsilon_{\beta} -U\langle
n_{\beta}\rangle+\hbar\omega_0+i\Gamma\left(\epsilon
+\hbar\omega_0\right)}\label{greenfba} \, ,
\end{eqnarray}
\begin{eqnarray}
G_{\alpha\beta}(\epsilon)&=&\frac{\lambda\left[\langle
b^{\dagger}b\rangle +1-\langle
n_{\alpha}\rangle\right]G_{\beta\beta}(\epsilon)}{\epsilon
-\epsilon_{\alpha}-U\langle
n_{\alpha}\rangle-\hbar\omega_0+i\Gamma\left(\epsilon-
\hbar\omega_0\right)} \label{greenfab} \, ,
\end{eqnarray}
where the expressions for the corresponding self-energies are
given by
\begin{eqnarray}
\Sigma_{\alpha}&=&\lambda^2\frac{\left[\langle
b^{\dagger}b\rangle+\langle n_{\beta}\rangle\right]\left(
\epsilon-\epsilon_{\alpha}-U\right)}{\epsilon
-\epsilon_{\beta}-U\langle n_{\beta}\rangle+\hbar\omega_0
+i\Gamma(\epsilon+\hbar\omega_0)},\\  \label{sigmaa}
\Sigma_{\beta}&=&\lambda^2\frac{\left[\langle
b^{\dagger}b\rangle+1-\langle n_{\alpha}\rangle\right]\left(
\epsilon-\epsilon_{\beta}-U\right)}{\epsilon
-\epsilon_{\alpha}-U\langle
n_{\alpha}\rangle-\hbar\omega_0+i\Gamma(\epsilon-\hbar\omega_0)}
\, . \label{sigmab}
\end{eqnarray}
We have considered here the paramagnetic case
with $\langle n_{i\bar\sigma}\rangle=\langle %
n_{i\sigma}\rangle\equiv\langle n_{i}\rangle$. The broadening due
to the leads is given by $\Gamma_i(\epsilon)=2\pi %
t^2_{i}\rho(\epsilon)$, where $\rho (\epsilon)$ is the DOS in the leads.
\cite{gen_QD_ref} We take $\Gamma_\alpha = \Gamma_\beta = \Gamma (\epsilon)$, for
simplicity. \footnote{$U$-dependent vertex renormalization of $\Gamma_i$ [e.g., A. Levy
Yeyati, A. Mart\'{\i}n-Rodero, and F. Flores, Lect. Notes in Physics {\bf 547}, 27
(2000)] may introduce minor effects. We tested implementation by W. Meztner [\prb {\bf
43}, 8549 (1991)] and found no appreciable changes of the results shown here (for
$\Delta\epsilon\gg\Gamma$).}

The nonlocal Green's functions, Eqs.\ (\ref{greenfba}) and
(\ref{greenfab}), have simple physical interpretations. At low
temperature $T$, $\langle b^{\dagger}b\rangle\approx 0$ thus
$G_{\beta\alpha}\rightarrow 0$ as $\langle
n_{\beta}\rangle\rightarrow 0$, i.e., there is no phonon
emission if the $\beta$ level is empty. It is also interesting to
note that if the dot $\alpha$ is not completely full, $\langle
n_{\alpha}\rangle <1$, the process of phonon absorption described
by $G_{\alpha\beta}$ is possible (even at small $T$).

In the limit $U=0$ the local Green's functions $G_{ii}$ have two
poles, one at $\epsilon_i$ and another at the pole of the
self-energy $\Sigma_{i}$. In this case there is a single phonon
resonance condition for both levels ($i=\alpha,\beta$), achieved
when the phonon energy $\hbar\omega_0$ matches the energy
difference between the two localized electron energies
$\Delta\epsilon\equiv\epsilon_{\beta}- \epsilon_{\alpha}$, 
i.e., at $\Delta\epsilon=\hbar\omega_0$. Thus, the resonance
effectively couples both sites.  Such a phonon resonance condition
on transport has been recently explored by Tasai and Eto.
\cite{eto} They find a sharp dip in the conductance as the result
of destructive interference between bonding and antibonding
states (see the dashed line in Fig.\ 2). We will see below that
electron repulsion greatly affects this behavior.

In the limit $\lambda=0$ we recover the Hubbard I approximation.
\cite{hubbard1} As expected, each local Green's function has poles
at $\epsilon_i$ and $\epsilon_i+U$. In contrast, when $U\neq 0$,
we find different resonant conditions for $\alpha$ and $\beta$.
The Coulomb repulsion requires extra energy for electrons to
tunnel into a fully or partially occupied state. This extra energy
depends on the occupation fraction of the two sites, which can be
controlled by the gate voltage. The self-consistent charge of each
site is obtained by integrating the DOS, $\rho_{ii}=(-1/\pi){\tt
Im} G_{ii}(\epsilon)$, for different gate voltages. With
appropriate parameter values we find that the effects brought
about by the EPI and EEI are emphasized under resonant conditions.
Hereafter all energy quantities are given in units of the level
spacing $\Delta\epsilon$.

\begin{figure}
\centerline{\includegraphics[width=2.8in]{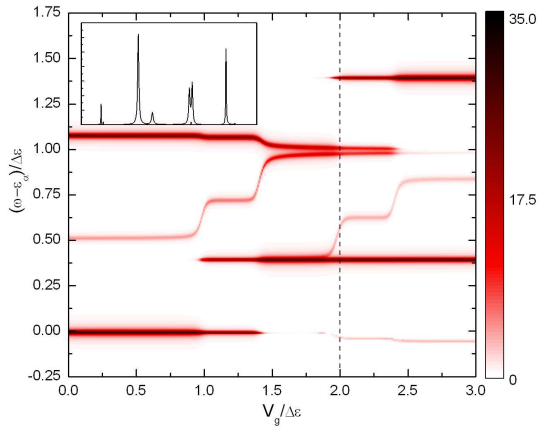}}
\caption{\label{fig2} (Color online) Total density of states (DOS)
vs energy and gate voltage for a two-site molecule as in Fig.\
\ref{Fig1}. In units of $\Delta\epsilon$, the parameters are:
$\lambda=0.2$, $k_{B}T=0.015$, $t_{\alpha}=t_{\beta}= 0.1$, $U =
0.4$, and $\hbar \omega_0 = 0.6$. The Fermi level is set at zero
energy; $\epsilon_{\alpha}=1$ and $\epsilon_{\beta}=2$ for
$V_g=0$. For $V_g \approx 1.6$--2.4, the DOS exhibits a clearly
split feature at $\omega - \epsilon_\alpha \approx 1$ produced by
the strong phonon-assisted level coupling. The inset shows the DOS at
$V_g=2$, indicated by the dashed line in the main figure.}
\end{figure}

Figure \ref{fig2} shows the total (for both sites) DOS for the
molecular system as a color map for different values of gate
voltage $V_g$ and energy $\omega$, measured with respect to the
lowest level $\epsilon_\alpha$. The temperature is set at
$k_BT=0.015$, $U=0.4$, and $\hbar\omega_0 = 0.6$. [The DOS is
normalized to obtain the correct charge number for each dot.]  For
low-voltage values, the molecule is empty and the DOS shows two
main features at the noninteracting energies $\omega \simeq
\epsilon_\alpha, \epsilon_\beta$, slightly shifted and broadened
by the coupling to the phonons and leads. As the gate voltage
increases, the lowest level approaches the Fermi level, and as its
occupancy grows a unique feature in the DOS develops. This is shown
in Fig.\ \ref{fig2} at energy $\epsilon_\alpha + U \approx 0.4$
for $V_g \gtrsim 1$. Notice that at the same time, a weak
phonon-related feature (a ``phonon replica") appears at energy
$\simeq \epsilon_\alpha + \hbar \omega_0 = 0.6$, as can be seen
from the expression for the self-energy, Eq.\ (\ref{sigmab}).
[This feature persists weakly even at low $V_g$ at $\simeq 0.5$,
due to the finite $\lambda$ and $T$ values in Fig.\ \ref{fig2}.]

For $V_g \gtrsim 1.4$, the lowest level is almost fully occupied,
$\langle n_\alpha \rangle \simeq 1$, and the resonant
absorption/emission condition in Eq.\ (\ref{sigmab}) is achieved
when $\epsilon_{\alpha}+U\langle n_{\alpha}\rangle+\hbar\omega_0
\simeq \epsilon_{\beta}$. The increasing $\langle n_\alpha
\rangle$ occupation shifts the phonon replica feature, moving it
closer to $\epsilon_\beta$. The resonant condition results in the
effective mixing of the $\epsilon_\alpha + U$ and $\epsilon_\beta$
levels due to phonon-assisted transitions. The mixing produces a
near degeneracy and, as a consequence, an effective phonon Rabi
splitting of spectral features appears. This is the origin of the
doublet appearing at $\omega - \epsilon_\alpha \simeq 1$ with
nearly equal-size peaks in the range $V_g \approx 1.6$--$2.4$. Note
that the lower level participates in the process as a result of
the Coulomb repulsion between electrons. Notice further that the
resonant condition disappears once the higher level becomes fully
(doubly) occupied ($V_g \gtrsim 2.4$ in Fig.\ \ref{fig2}). The DOS
returns then to the standard Hubbard peaks at $\epsilon_\alpha +U$
and $\epsilon_\beta +U$ dominating the spectrum of the molecule
(weak phonon-related features near 0 and 0.8 are present for
finite $T$ and $\lambda$). As we will show below, the Rabi
resonance in the presence of EEI affects also the conductance
through the system, providing an experimentally accessible
signature of the effect.

\begin{figure}
\centerline{\includegraphics[width=2.8in]{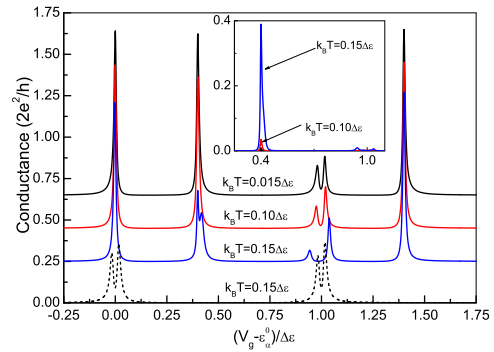}}
 \caption{\label{Fig3} (Color online) Conductance vs gate voltage
for $k_BT/\Delta\epsilon=0.015$, 0.10, and 0.15. Temperature has
a strong effect on the resonance condition, enhancing the Rabi
splitting of the third Coulomb blockade peak ($V_g -
\epsilon_\alpha \simeq 1$) at higher $T$ and exposing the shoulder in
second peak. The inset shows inelastic contributions for the
conductance. The highest peak is due to phonon emission. The dashed
line shows results for $U=0$ and $\omega_0=1$.  Curves offset for
clarity. }
\end{figure}

\begin{figure}[t]
\centerline{\includegraphics[width=2.8in]{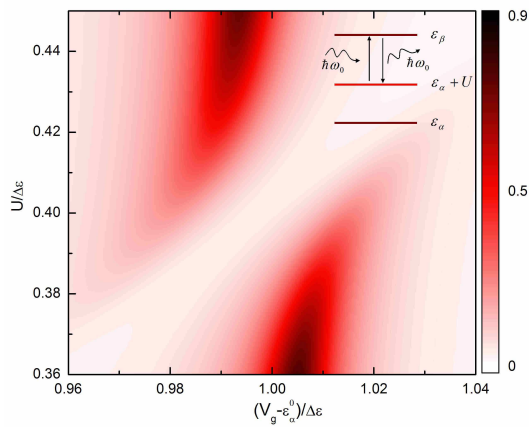}}
\caption{\label{Fig4} (Color online) Conductance through the
molecule at around the third Coulomb blockade peak as a function of
interaction $U$ and gate voltage $V_g$ for
$k_BT=0.015\Delta\epsilon$. The inset indicates phonon emission and
absorption processes involved.}
\end{figure}

Figure \ref{Fig3} shows the conductance $G$ vs gate voltage for
different temperatures ($G\propto|G_{\bar 11}|^2$ where $G_{\bar
11}\propto
t^2_{\alpha}G_{\alpha\alpha}+t^2_{\beta}G_{\beta\beta}+t_{\alpha}
[G_{\alpha\beta}+G_{\beta\alpha}] t_{\beta}$). As the gate voltage
increases, the conductance exhibits the anticipated Coulomb
blockade (CB) peaks. Notice that the first two are
associated here with $\epsilon_\alpha$, and show nearly full
$e^2/h$ conductance (limited by finite temperatures).  The
dominant effect of higher temperatures is to produce important
changes in the dot occupancy due to phonon emission (occupancy
changes due to thermal excitation of electrons are negligible).
The third CB peak appearing at $V_g - \epsilon_\alpha ^0 \simeq 1$
is clearly split, indicating the phonon-mediated transitions
between the $\alpha$ and $\beta$ levels. The temperature
dependence of this peak is pronounced, as the thermal variation in
electron and phonon occupations will not only make it weaker but
will also rapidly detune the resonance condition. A higher phonon
presence results in stronger EPI and {\em a drop} in the
conductance, a typical signature of {\em interference} between
resonant tunneling and polaron formation processes.  The splitting
of the third CB peak increases {\em linearly} with the EPI
strength $\lambda$, as expected in Rabi splitting phenomena.
Notice that once the $\beta$ level is nearly full, the effect
disappears, and the Coulomb blockade peak at $V_g -
\epsilon_\alpha^0 \simeq U + \Delta \epsilon = 1.4$ is essentially
$T$ independent. The appearance of a shoulder in the {\em second}
CB peak for higher temperatures is also a consequence of EPI, but
one that is inelastic in nature: The inset of Fig.\ \ref{Fig3}
shows the contribution of {\em inelastic} processes ($\propto
G_{\alpha\beta} + G_{\beta\alpha}$) to the conductance. The
highest peak at $V_g - \epsilon_\alpha^0 \simeq 0.4$ is due to
phonon emission processes: electrons thermally excited to level
$\epsilon_{\beta}$ (above the Fermi level here) fall to level
$\epsilon_{\alpha}+U$ by the emission of phonons (the inset of Fig.\
\ref{Fig4} illustrates levels involved). These processes do {\em
not} enhance the conductance of the resonant level, but rather
{\em reduce} it, as one sees in the main panel. The suppression is
produced by the destructive interference of the two different
conducting processes that electrons undergo.

The resonant condition of phonon-assisted tunneling through the
molecule in the presence of EEI is a function of the interaction
parameter $U$ and the occupation of the molecule. In order to
explore the dependence of the Rabi splitting on interaction
parameters we show in Fig.\ \ref{Fig4} a color map of the
conductance as function of $U/\Delta\epsilon$ and
$(V_g-\epsilon^0_{\alpha})/\Delta\epsilon$. It is clear that the
conductance returns to the non-phonon-assisted resonant single CB
peak whenever $U/\Delta \epsilon$ is far below or above the
resonant value $U \simeq \Delta \epsilon - \hbar \omega_0 =0.4$.
The lowest level is nearly doubly occupied in this $V_g$ regime,
i.e., $(V_g-\epsilon^0_{\alpha})/\Delta\epsilon \simeq 1$. At
the resonance, $U\simeq 0.4$, the conductance Rabisplits into two
peaks that have the same height and demonstrates the effect of
polaron formation on the conductance.

We have shown that the competition of EEI and EPI in a diatomic
molecule produces unexpected Rabi splitting phenomena in the DOS
with observable effects on the conductance. This phenomenon
involves states produced by the Coulomb repulsion, and it is
enhanced at higher temperatures, a direct consequence of the
thermal nature of the phonon bath involved.

We would like to thank CAPES (Brazil) and NSF-IMC grant No 0336431
for support, and C. B\"usser for helpful discussions.


\begin{thebibliography}{apssamp}
\bibitem{f1} Electron correlations with the leads are not considered
in the present work. Thus, the physics described is in the regime
$T>T_K$, where $T_K$ is the Kondo temperature.
\end{thebibliography}
\end{document}